\documentclass[a4paper,11pt]{article}
\usepackage{float}
\usepackage{jinstpub}
\usepackage{graphicx}
\usepackage{subfigure}
\usepackage{xcolor}

\usepackage{multirow}
\usepackage{amsmath}
\usepackage[numbers,sort&compress]{natbib}
\usepackage{verbatim}
\usepackage{textgreek}
\usepackage{ulem}
\usepackage{caption}
\author[a]{Junyou Chen,}
\author[b,1]{Jilei Xu,\note{Corresponding author}}
\author[a,2]{Yongbo Huang,\note{Corresponding author}}
\author[b]{Sibo Wang,}
\author[b,c]{Chuanshi Dong,}
\author[b]{Haoqi Lu,}
\author[b,c]{Changgen Yang,}
\author[b]{Yongpeng Zhang,}
\author[b]{Yi Wang}

\affiliation[a]{School of Physical Science and Technology, Guangxi University, Nanning 530004, China.}
\affiliation[b]{Institute of High Energy Physics, Beijing, 100049, China}
\affiliation[c]{University of Chinese Academy of Sciences, Beijing, 100049, China}

\emailAdd{xujl@ihep.ac.cn}
\emailAdd{huangyb@gxu.edu.cn}

\title{Study of a Compact Device for Water Attenuation Length Measurements}
\date{ }

\abstract{ 
This study presents the development and validation of a compact device for measuring the water attenuation length (WAL), utilizing photomultiplier tubes (PMTs), optical fibers, and light-emitting diodes (LEDs). An 8~m water tank and the device was constructed and validated in the laboratory. The device is capable of measuring WAL values up to 50~m. The stray light was blocked mainly by a custom-designed shutter. Toy Monte Carlo simulations were employed to evaluate the measurement uncertainty, which was found to be within reasonable limits. These simulations further indicate that the uncertainty can be reduced and more accurately predicted for a larger-scale device with a length of 30~m. Real-time monitoring was achieved by integrating the device into a water purification circulation system, providing a practical, scalable solution for WAL measurement in future large-scale water Cherenkov detectors.}


\begin{document}
\maketitle
\flushbottom

\section{Introduction}\label{sec1}
Water Cherenkov detectors have become indispensable tools for detecting elementary particles in high-energy physics. With the advancement of neutrino and primary cosmic ray detection, these detectors are increasing in size, leading to increasingly stringent water quality requirements. The water attenuation length (WAL) is a critical parameter for assessing water quality, and its measurement is essential to ensure the stable and efficient operation of water Cherenkov detectors. As summarized in~\cite{design-proposal}, two primary methods exist for measuring WAL: offline laboratory measurements and online measurements integrated within the detector.

For offline laboratory measurements, experiments such as Daya Bay~\cite{DayaBay}, LHAASO~\cite{LHAASO0,LHAASO1,LHAASO2} have detected WAL values in 30~m using LEDs and photomultiplier tubes (PMT). CHIPS~\cite{CHIPS} utilized a laser and a photodiode to measure a WAL of approximately 139~m, while UDEAL~\cite{UDEAL0,UDEAL1} employed lasers and integrating spheres to achieve a WAL measurement of around 80~m. For online measurement, Super-K~\cite{Super-K} have provided precise results, achieving a WAL of approximately 97.9~m using an LED and a diffuser ball. TRIDENT~\cite{TRIDENT} and ANTARES~\cite{ANTARES} utilized diffusers and photosensors to measure WAL values ranging from of 20~m to 60~m in the deep sea.
 
A novel design for a WAL measurement device integrated within a large water Cherenkov detector was proposed in paper~\cite{design-proposal}, offering advantages such as operational stability, low cost, and online measurement capability. Toy Monte Carlo (ToyMC) simulations demonstrated the device's potential for effective performance in a 100-meter range with a device length 30~m. To validate this design, an 8~m prototype was constructed and tested in the laboratory.

This work presents the first practical implementation and experimental validation of this compact device for WAL measurement. The experimental setup is detailed in Section~\ref{sec2}; the optimization procedures are described in Section \ref{sec3}; the WAL measurement and analysis are presented in Section \ref{sec4}; and conclusions are given in Section~\ref{sec5}.

\section{Experimental setup}
\label{sec2}

The basic idea is to use a point-like source to illuminate PMTs at the same time. The PMT charges obey the inverse-square law, $d^{-2}$, in the air. If in the water, an additional exponential term has to be added. The formula is expressed as follows: 

\begin{equation}
    Q(d)\sim d^{-2}e^{-d/{\lambda}}
    \label{eq:principle}
\end{equation}

Where $d$ is the PMT distance to the light source and $\lambda$ is the WAL. Based on this, a point-like source and eight PMTs were constructed and integrated into a 8~m water tank.

\begin{figure}[h]
  \centering
  \includegraphics[width=0.6\textwidth]{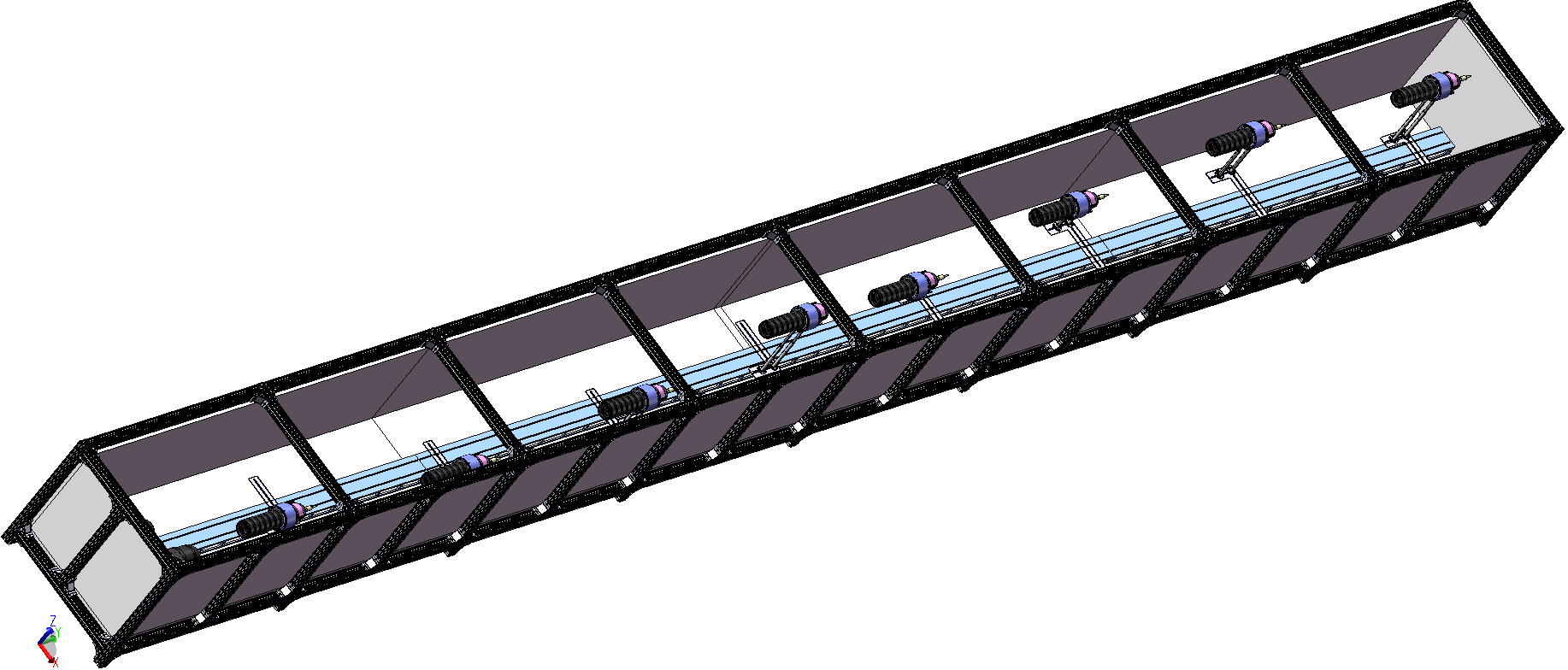}
  \caption{The designed structure of water tank with size of 8 m × 0.6 m × 0.5 m, and eight PMTs were placed in it. }
  \label{fig:waterTank2}
\end{figure}

\begin{figure*}[htbp]
        \centering
        \subfigure[]{
            \includegraphics[width=0.4\hsize]{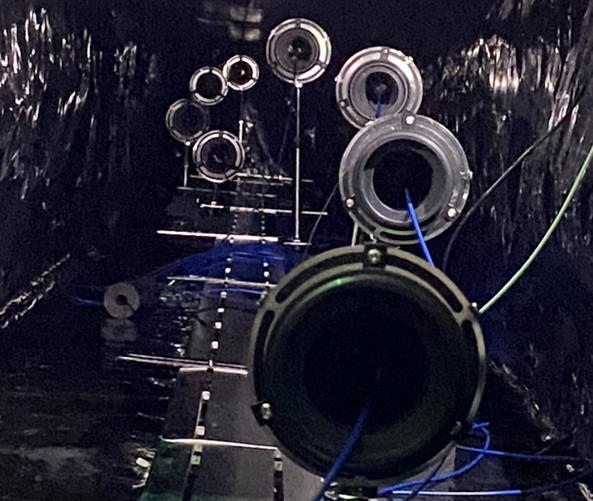}
        	\label{fig:hdpe}
        }
        \quad
        \subfigure[]{
            \includegraphics[width=0.4\hsize]{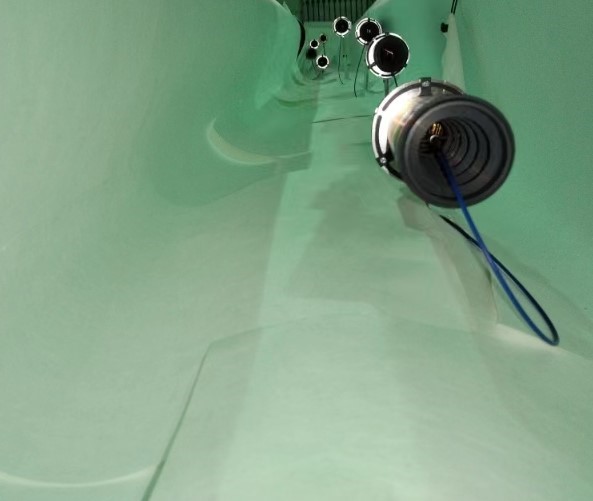}
        	\label{fig:tyvek}
        }
        \caption{Internal view of the device under the (a) HDPE configuration and (b) Tyvek configuration. The PMTs with their aligned shutters are also visible.} 
    	\label{fig:hdpe_tyvek}
\end{figure*}

The water tank dimensions were 8 m $\times$ 0.6 m $\times$ 0.5 m, and the tank was built with an aluminum profile frame, lined internally with solid wood panels, providing a capacity of up to two tons of ultra-pure water, as shown in Fig.~\ref{fig:waterTank2}. To prevent water leakage, the entire inner surface was lined with a large sheet of high-density polyethylene (HDPE) plastic film with a thickness of 600~$\mu$m. The top edges of the HDPE film were extended and overlapped to ensure light tightness. To study the impact of different environmental conditions and light reflection, two distinct configurations were applied to the tank's inner wall during WAL measurements: a black HDPE film and a white Tyvek film, as shown in Fig.~\ref{fig:hdpe_tyvek}.

The black HDPE was installed on the inner walls to minimize reflectivity and reduce stray light interference. However, during the WAL study, we observed that the smooth surface of the black HDPE could still induce mirror-like reflections at large incident angles, contributing to stray light. Consequently, the HDPE surface was roughened with low-grit sandpaper to mitigate specular reflection. Furthermore, HDPE effectively prevents water leakage and ensures stable water quality inside the tank~\cite{HDPE}. For the Tyvek configuration, the inner wall was lined with a white ultra-fine high-density polyethylene fiber material~\cite{Tyvek}, which is highly reflective. Large-scale neutrino experiments, such as JUNO and Super-K, often employ Tyvek film on their detector walls to enhance Cherenkov light collection.

\begin{figure*}[htbp]
        \centering
        \subfigure[]{
            \includegraphics[width=0.55\hsize]{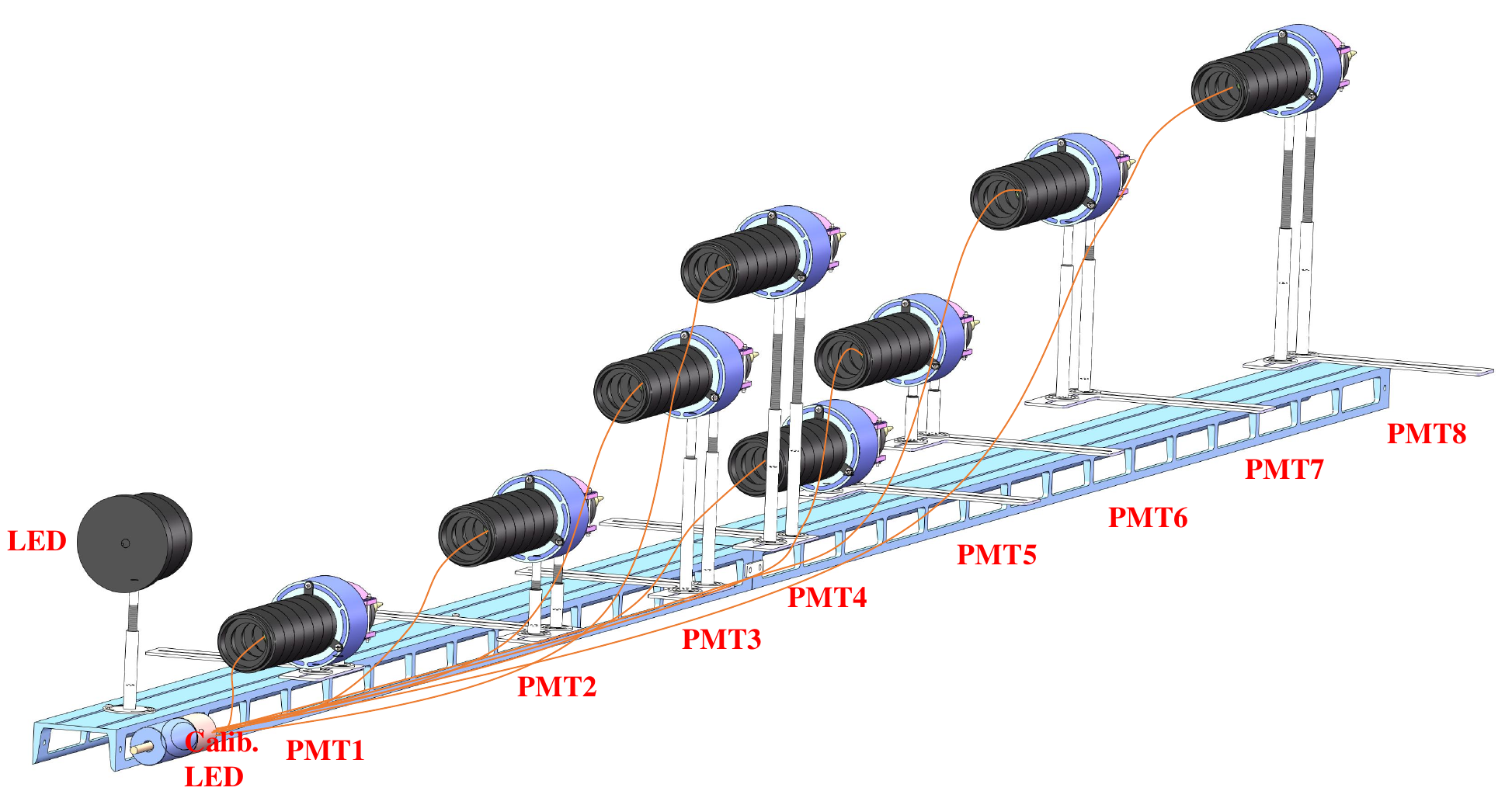}
        	\label{fig:PMT Axial}
        }
        \quad
        \subfigure[]{
            \includegraphics[width=0.25\hsize]{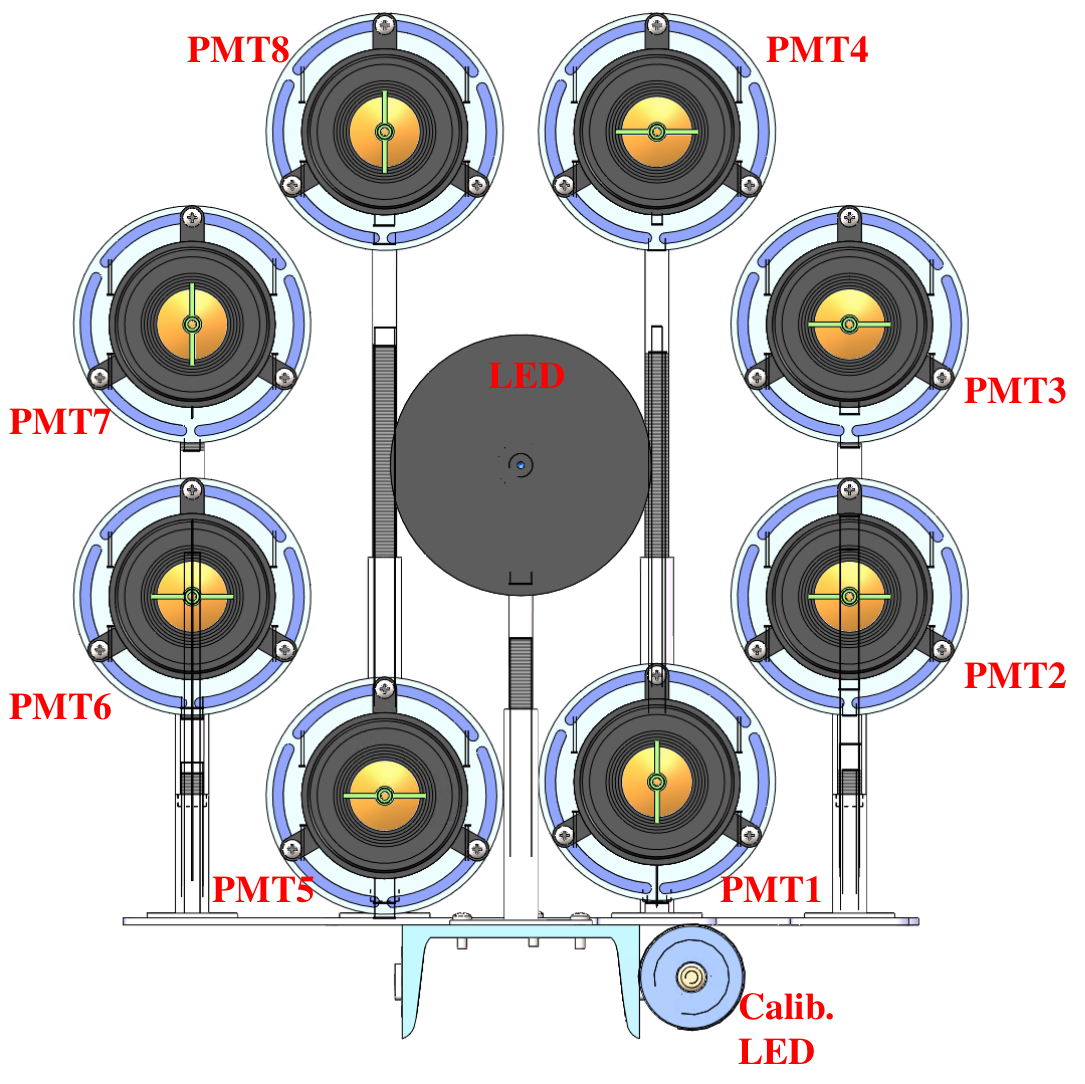}
        	\label{fig:PMT lateral}
        }
        \caption{(a) 3-dimensional view and (b) cross-sectional view of the core component of the WAL measurement device.} 
    	\label{fig:Axial_lateral}
\end{figure*}

\begin{figure}[htbp]
        \centering

            \includegraphics[width=0.45\hsize]{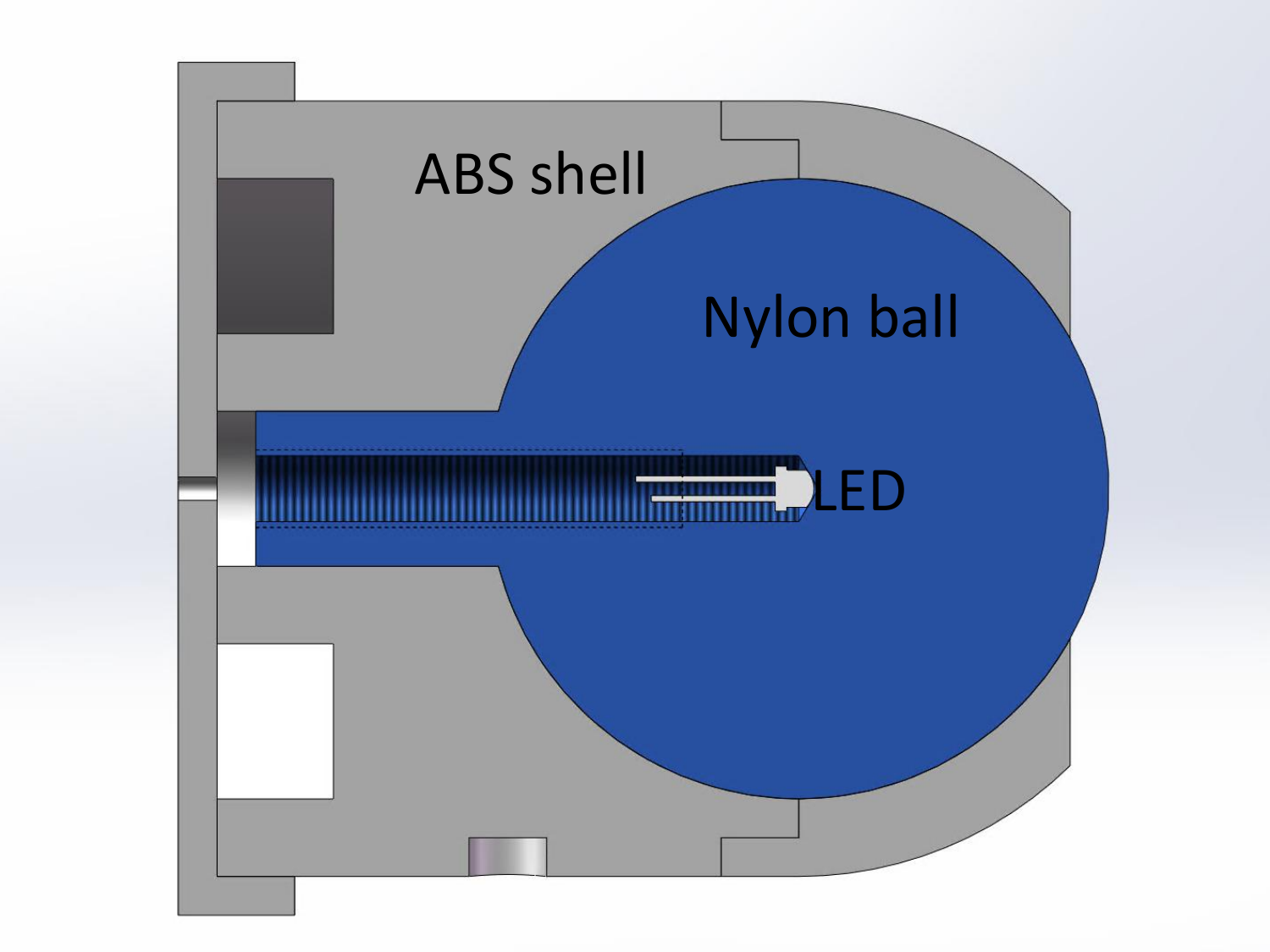}

        \caption{Side cross-sectional view of the diffuser. A single LED is positioned at the center of a nylon ball. An ABS shell, composed of three parts, is fastened together with screws.} 
    	\label{fig:Diffuse_ball}
\end{figure}

The measurement device integrated into the water tank consists of LEDs with wavelength 400~nm~\cite{ledshop}, diffuser balls, optical quartz fibers~\cite{fiber-ref}, and 3-inch dynode PMTs produced by HZC~\cite{HNZC-3-inch-PMT,PMTs-perform-optimally}. Eight PMTs were symmetrically placed around the LED light source, each positioned at a different distance and numbered sequentially from the closest to the farthest, with an interval of approximately 1~m, as shown in Fig.~\ref{fig:Axial_lateral}. One LED was inserted into a nylon ball (model 66, 8 cm in diameter) serving as a diffuser, as illustrated in Fig.~\ref{fig:Diffuse_ball}. The diffuser was encased in a black Acrylonitrile Butadiene Styrene (ABS) shell, with only a small 1 cm diameter aperture left to function as a point-like light source. The non-uniformity of the diffuser has been studied in \cite{design-proposal}. The diffuser and the eight PMTs were mounted on a stainless steel rail to ensure mechanical robustness and prevent relative displacement in case of minor tank deformation. 

\begin{figure}[htbp]
        \centering

            \includegraphics[width=0.5\hsize]{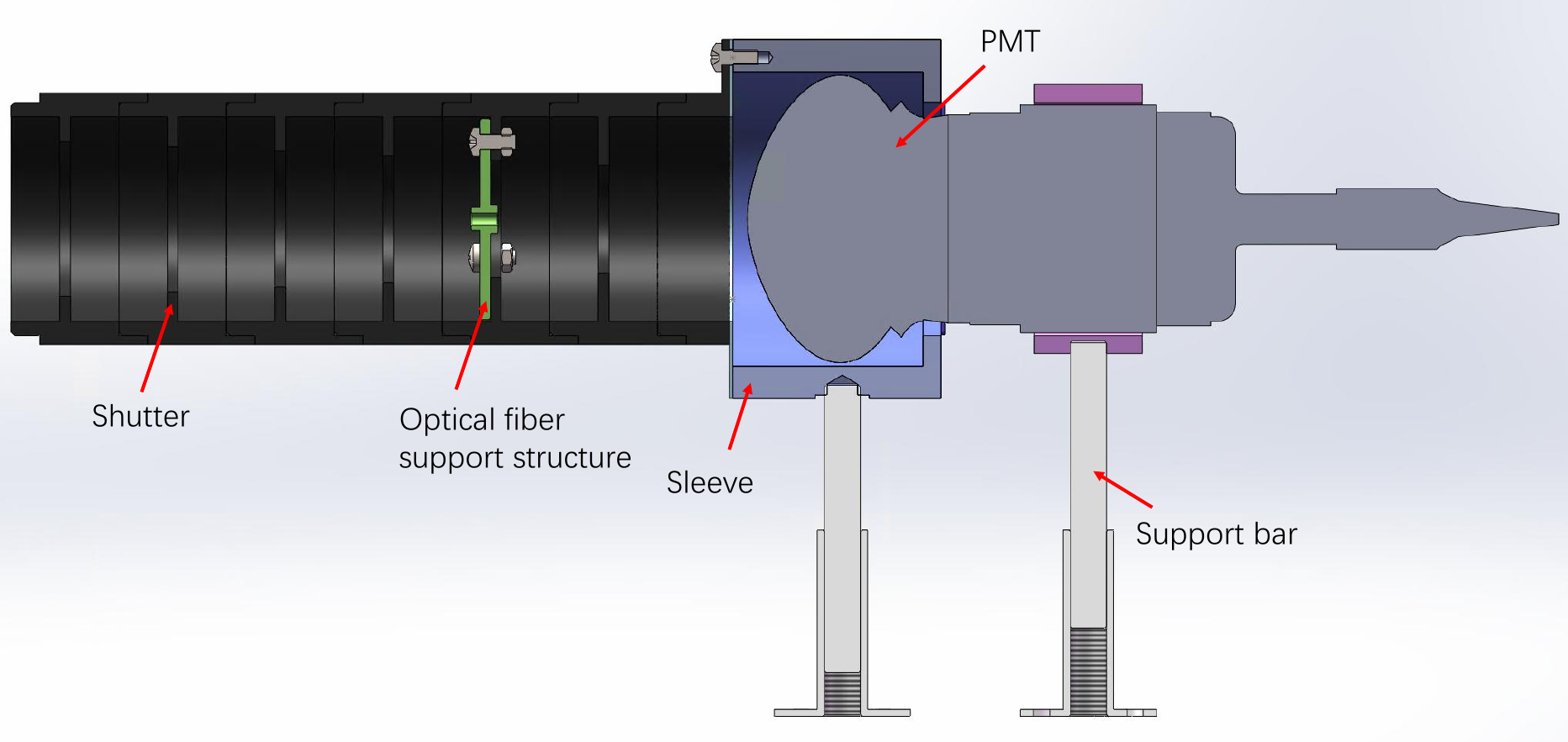}

        \caption{Side cross-sectional view of a single PMT with its light guide and support bars. The green component in the shutter represents the optical fiber support structure.} 
    	\label{fig:PMT1_PMTb}
\end{figure}

Each PMT was housed within an ABS plastic sleeve. A light guide, assembled from six mechanically joined shutters, was mounted in front of each photocathode, as shown in Fig.~\ref{fig:PMT1_PMTb}. Each guide was oriented towards the LED to ensure direct light incidence on the photocathode and to shield against stray light. Each PMT was supported by two bars, which could be adjusted to precisely align the guide's direction. Eight optical fibers, each with an identical length of 4~m and a diameter of 200~$\mu$m, were used. One end of each fiber was connected to a small nylon ball (3~cm in diameter) coupled to the calibration LED, and the other end was fixed within the shutter of its respective PMT \cite{fiber-ref}.

For long-term WAL monitoring, the water tank was connected to a water circulation and purification system~\cite{circulation-system}. Data acquisition from the eight PMTs was performed using a Flash ADC~\cite{haiqiong-paper}. After water purification was completed, the circulation was halted to conduct long-term online monitoring.
 
\section{Stray light shielding strategy}
\label{sec3}

While stray light enhances light collection and detection efficiency in large-scale neutrino experiments, justifying their highly reflective environments, it adversely affects the accuracy of our measurement device. Stray light introduces additional photons to the PMTs, resulting in measurement uncertainties. Therefore, effective shielding against stray light is crucial for obtaining precise and reliable measurements. In this study, the system shielding effectiveness is evaluated by verifying the inverse-square law, and the power function for fitting is expressed as follows:

\begin{equation}
    Q(d)=Ad^{B}
    \label{eq:attL1}
\end{equation}

where $Q(d)$ denotes the charge collected by the PMT at a distance $d$ from the light source, $A$ is a constant coefficient, and $B$ is the exponent. Given that the attenuation of visible light from the LED point source in air is negligible under the experimental conditions, the value of $B$ is expected to be approximately $-2$.

\begin{figure}[htbp]
        \centering

            \includegraphics[width=0.45\hsize]{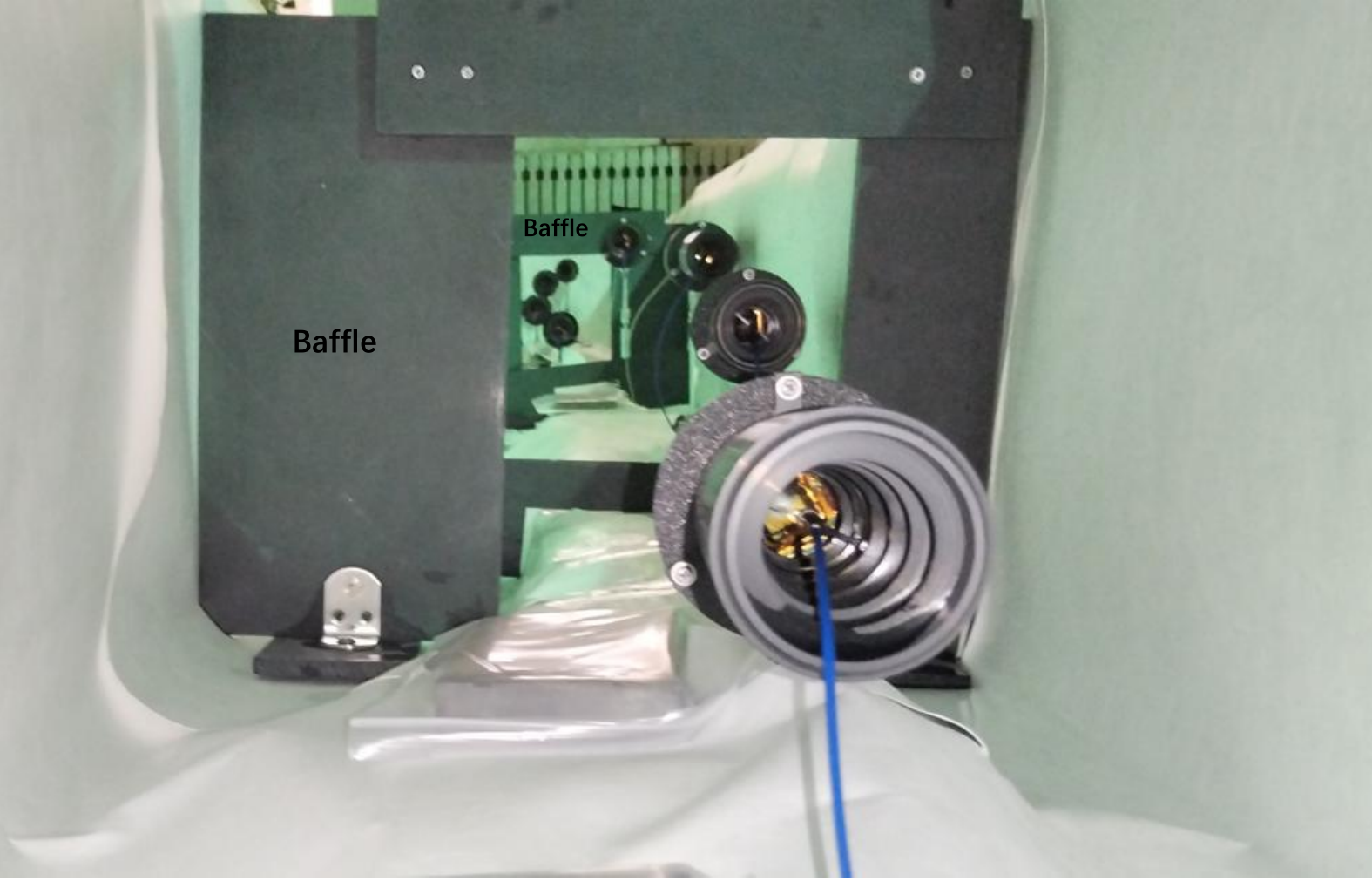}

        \caption{Two baffles are placed behind the PMT1 and PMT4 to reduce stray light.} 
    	\label{fig:baffle}
\end{figure}

\begin{figure}[htbp]
        \centering
        \subfigure[]{
            \includegraphics[width=0.4\hsize]{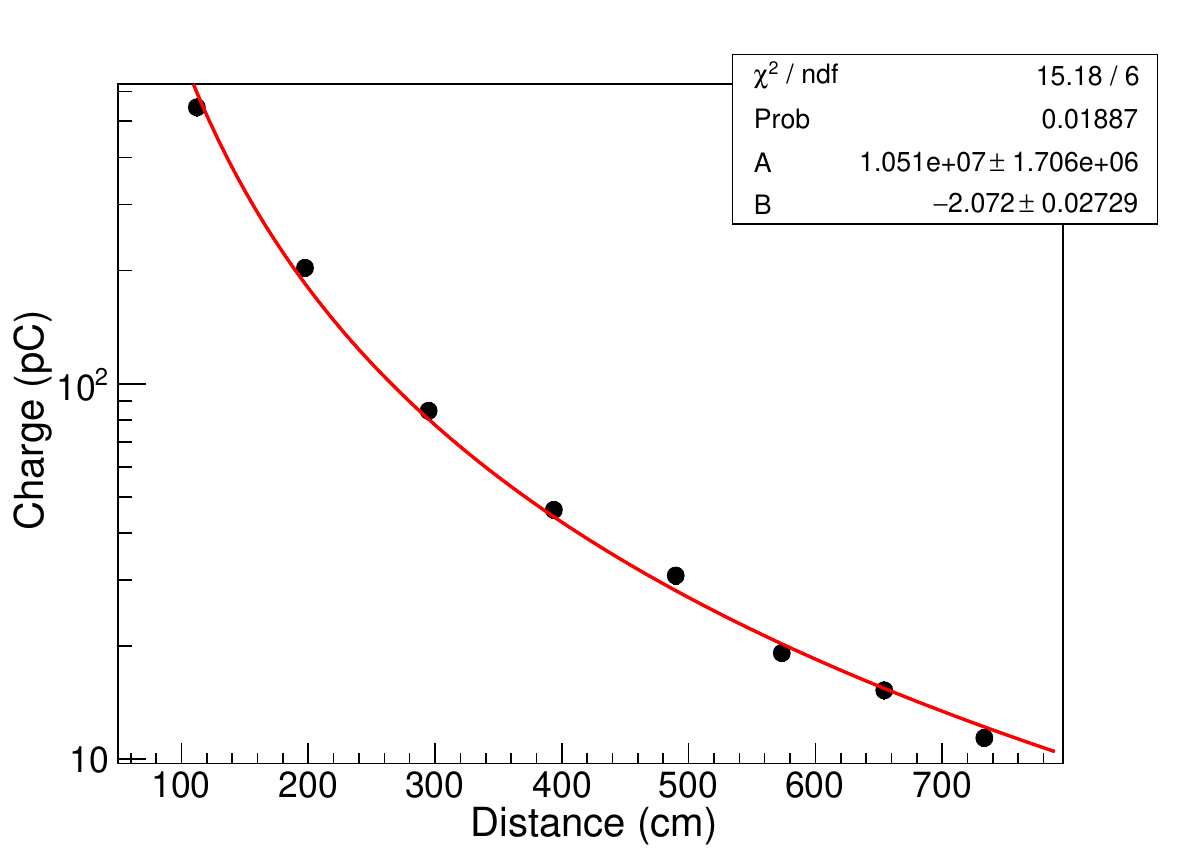}
        	\label{fig:tyvekair}
        }
        \quad
        \subfigure[]{
            \includegraphics[width=0.4\hsize]{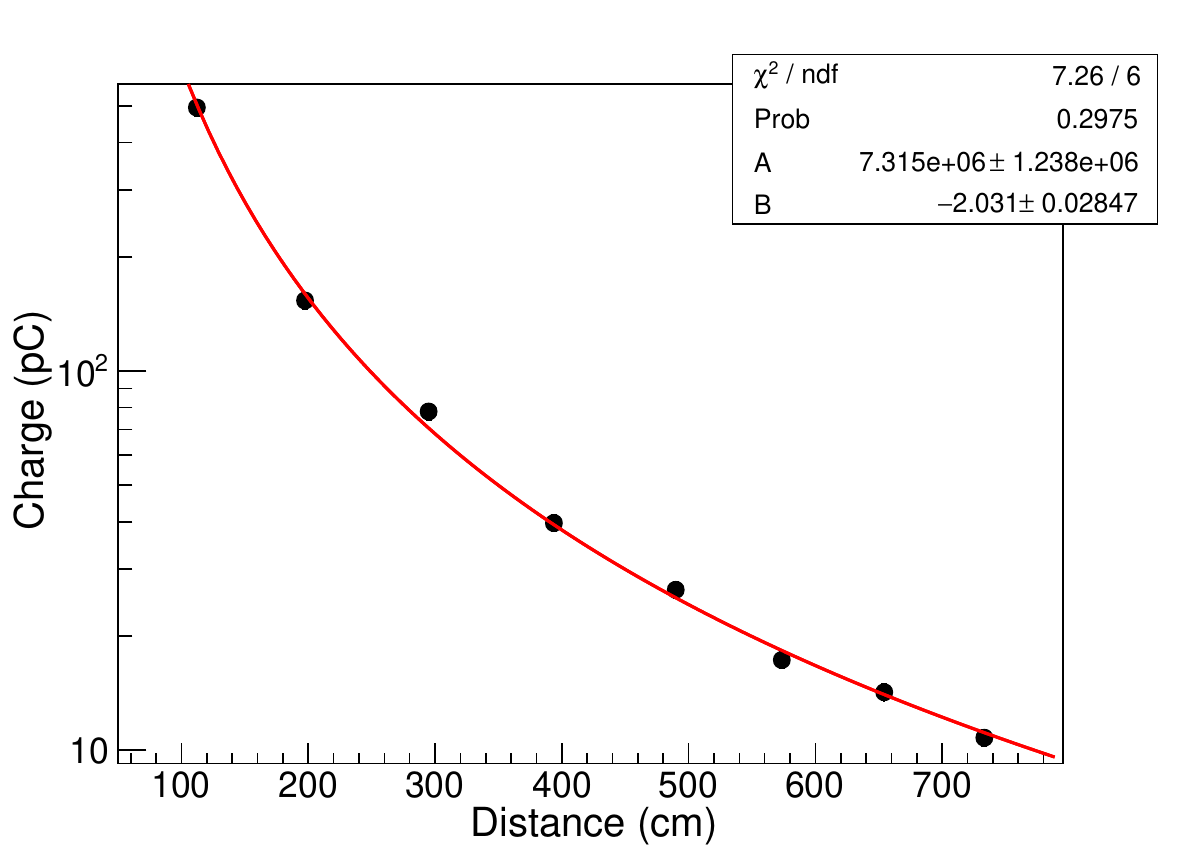}
        	\label{fig:hdpeair}
        }
        \caption{
        Verification of the inverse square law for point light sources in (a) Tyvek and (b) HDPE configurations with baffles. The black dots represent the charge collected by each PMT at the different distances from the point-like light source.} 
    	\label{fig:hdpeair_tyvekair}
\end{figure}

Prior to measurements, extensive preparatory work was conducted. This included calibrating the PMTs, evaluating the uniformity of the LED diffuser sphere and optical fibers, and characterizing PMT nonlinearity under varying light intensities. A critical step involved calibrating channel-to-channel deviations among the PMTs and associated electronics, constraining these differences to within $\pm 3\%$. Further details regarding these preparatory procedures are documented in~\cite{design-proposal,Huang-2017abb,Tang-2024jfs}. Subsequent verification of the inverse-square law for point sources was conducted in a light-tight tank lined with either HDPE or Tyvek. The smooth HDPE liner produced mirror-like reflections, resulting in a fitted power-law exponent $B = -1.56$, which deviated significantly from the ideal value of $-2$. To suppress these specular reflections, the HDPE film was mechanically roughened with sandpaper~\cite{sandpaper} and reinstalled. This modification yielded an improved fit of $B = -1.96$. For the Tyvek-lined tank, the high wall reflectivity increased the stray light component, resulting in $B = -1.82$. A baffle was intalled in the light-convergence region, as shown in Fig.~\ref{fig:baffle}, which reduced stray light, improving the fit to $B = -2.07$ as shown in Fig.~\ref{fig:tyvekair}. Applying the same baffle to the tank with the roughened HDPE liner further refined the result from $B = -1.96$ to $B = -2.03$, as shown in Fig.~\ref{fig:hdpeair}. These optimizations effectively suppressed stray light and recovered the expected inverse-square behavior, fulfilling the baseline requirements for subsequent WAL measurements.

\section{Measurement scheme and data analysis}
\label{sec4}
\subsection{Optimization of measurement scheme}
When light propagates through a transparent medium such as water, it obeys the Beer-Lambert law of exponential attenuation. Consequently, an exponential attenuation term is incorporated into Eq.~(\ref{eq:attL1}) as follows:

\begin{equation}
    Q_{i}(d_{i})=A d_{i}^{-B}e^{-d_{i}/{\lambda}}
    \label{eq:attL2}
\end{equation}

where $Q_{i}$ denotes the charge detected by the $i^{\text{th}}$ PMT at a distance $d_{i}$ from the LED source, $A$ is a constant coefficient, and $\lambda$ represents the light attenuation length in the medium.

It is evident that during measurement, the inverse-square effect of light propagation and its attenuation within the medium are coupled, potentially introducing systematic uncertainty. To decouple these effects, the measurement scheme was optimized by performing measurements in both air and water. The ratio between the two sets of measurements ($Q_{i}^{\rm water}$ and $Q_{i}^{\rm air}$) was then calculated, as expressed in Eq.~(\ref{eq:attL3}), to eliminate the potential impact of inaccuracies in the inverse-square term arising from experimental setup limitations. Given that the attenuation of visible light in air is negligible under the conditions of this experiment, the exponential attenuation term for air is omitted. This measurement scheme effectively cancels out influences caused by light intensity instability, differences in PMT characteristics (e.g., quantum efficiency, photocathode area), and variations among electronic channels, thereby reducing systematic errors and improving measurement reliability.

\begin{equation}
    Q_{i}^{\rm water}/Q_{i}^{\rm air}=Ce^{-d_{i}/{\lambda}}
    \label{eq:attL3}
\end{equation}

\subsection{HDPE configuration for WAL long-term observation}

When did the experiment, all eight PMTs were tuned to gain 3 $\times$ 10$^{6}$, and each channel's charge difference was calibrated by the calibration LED. The diffuser was lighted by pulse generator which produced pulse frequency 1~kHz and pulse width 100~ns. The light intensity was tuned to let the charge of first PMT (PMT1 in Fig.~\ref{fig:Axial_lateral}) obtained about 500 - 600 photoelectrons (PEs) in each pulse trigger, while the farthest PMT (PMT8) obtained about 8 - 10 PEs. Before filled the ultra-pure water in the tank, the eight PMT charges ($Q_{i}^{\rm air}$) were obtained firstly. The ultra-pure water with electrical resistivity 18~M$\Omega$$\cdot$cm was get from the ultra-pure water station and filled in a one-ton PE tank, and transported to the laboratory, pumped into the water tank by a clean vacuum pump. Then, the eight PMTs were triggered by LED diffsuer and charges ($Q_{i}^{\rm water}$) were measured by data acquisition system. 

\begin{figure}[htbp]
    \centering
    \subfigure[]{
        \includegraphics[width=0.4\hsize]{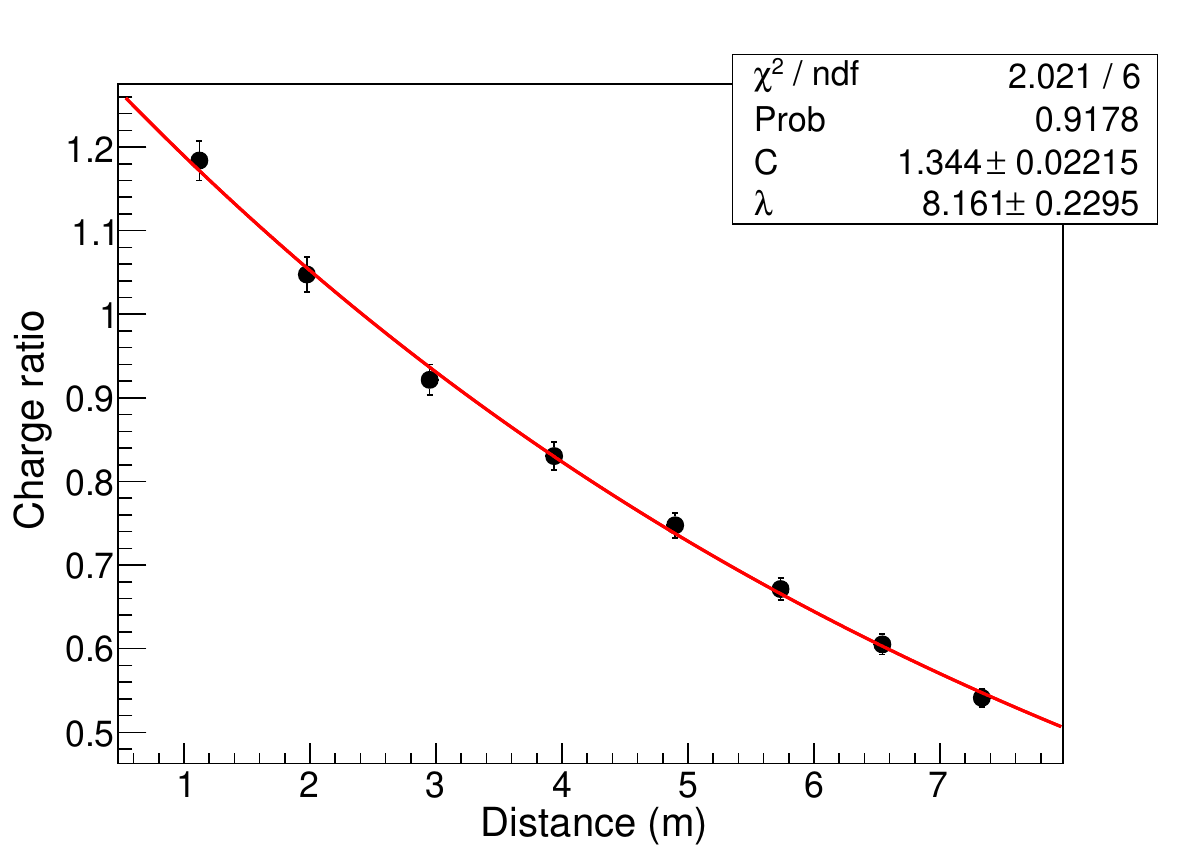}
        \label{fig:WAL}
    }
        \quad
    \subfigure[]{
        \includegraphics[width=0.4\hsize]{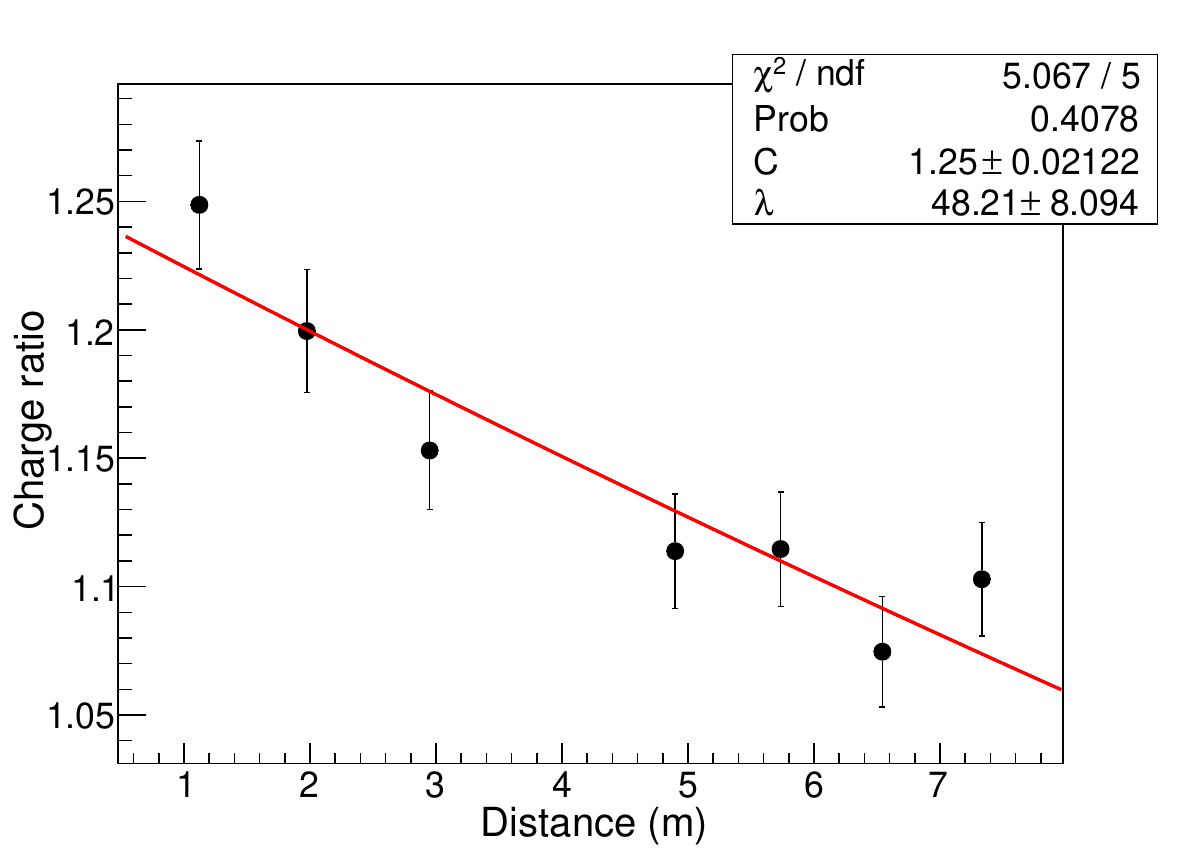}
        \label{fig:HDPE_initial_measurement}
    }
    \quad
    \subfigure[]{
        \includegraphics[width=0.4\hsize]{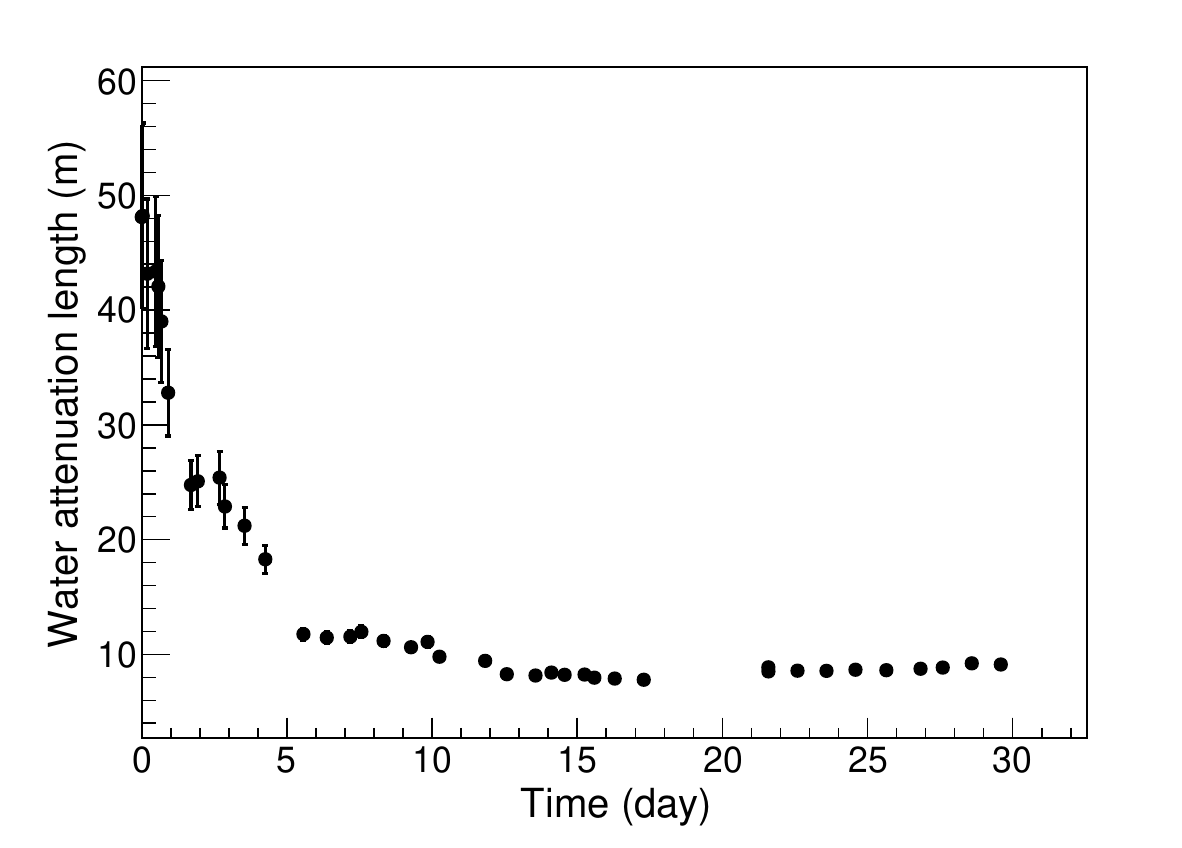}
        \label{fig:HDPE}
    }

    \caption{(a) An example of the WAL fitting with WAL 8.2~m based on Eq.~\ref{eq:attL3} during the long-term ultra-pure water monitoring. (b) Measurement result obtained immediately after ultra-pure water filling. (c) Long-term observation of the WAL in HDPE configuration. }
    \label{fig:WAL_and_HDPE}
\end{figure}

Finally, the data was processed, and the ratio of charge in water and air was fitted by Eq.~\ref{eq:attL3}. An example of WAL measurement in the HDPE configuration during the monitoring stage is shown in Fig.~\ref{fig:WAL}. The red curve represents the exponential fitting result, and the WAL ($\lambda$) is ($8.2 \pm 0.2$)~m. After initially filling the tank with pure water, an inspection revealed slight deformation in the black HDPE liner on the inner side of the lid near the fourth PMT. The data from seven PMTs were used for the Water Attenuation Length (WAL) fitting, as shown in Fig.~\ref{fig:HDPE_initial_measurement}. The initial WAL was estimated to be approximately 48.2~m, but with a relatively large uncertainty of 8.1~m. As the water purity increases and it becomes more transparent, the influence of stray light becomes more severe. This effect causes the photon collection of the PMTs to deviate from the ideal exponential attenuation model.

Real-time monitoring of the WAL was performed over a 30-day period within the HDPE-lined tank after filled the ultra-pure water, as shown in Fig.~\ref{fig:HDPE}. Usually, when the ultra-pure water produce by the ultra-pure water station, the WAL is more than 100~m, while after filled into the water tank, the WAL was reduced to 40 - 50~m. This was caused by the cleanness of the water tank, material impurity of the container, contact with air, and so on. Analysis of the long-term monitoring data indicated a significant decline in WAL during the first 10 days after filling, signifying progressive ultra-pure water quality deterioration, as shown in Fig.~\ref{fig:HDPE}. Subsequently, the WAL was stabilized and exhibited a slight upward trend, likely attributable to the settling of particulate matter, which marginally enhanced light transmission.

\subsection{Water quality purification and Tyvek configuration for long-term WAL observation.}

\begin{figure}[htbp]
        \centering
        \subfigure[]{
            \includegraphics[width=0.4\hsize]{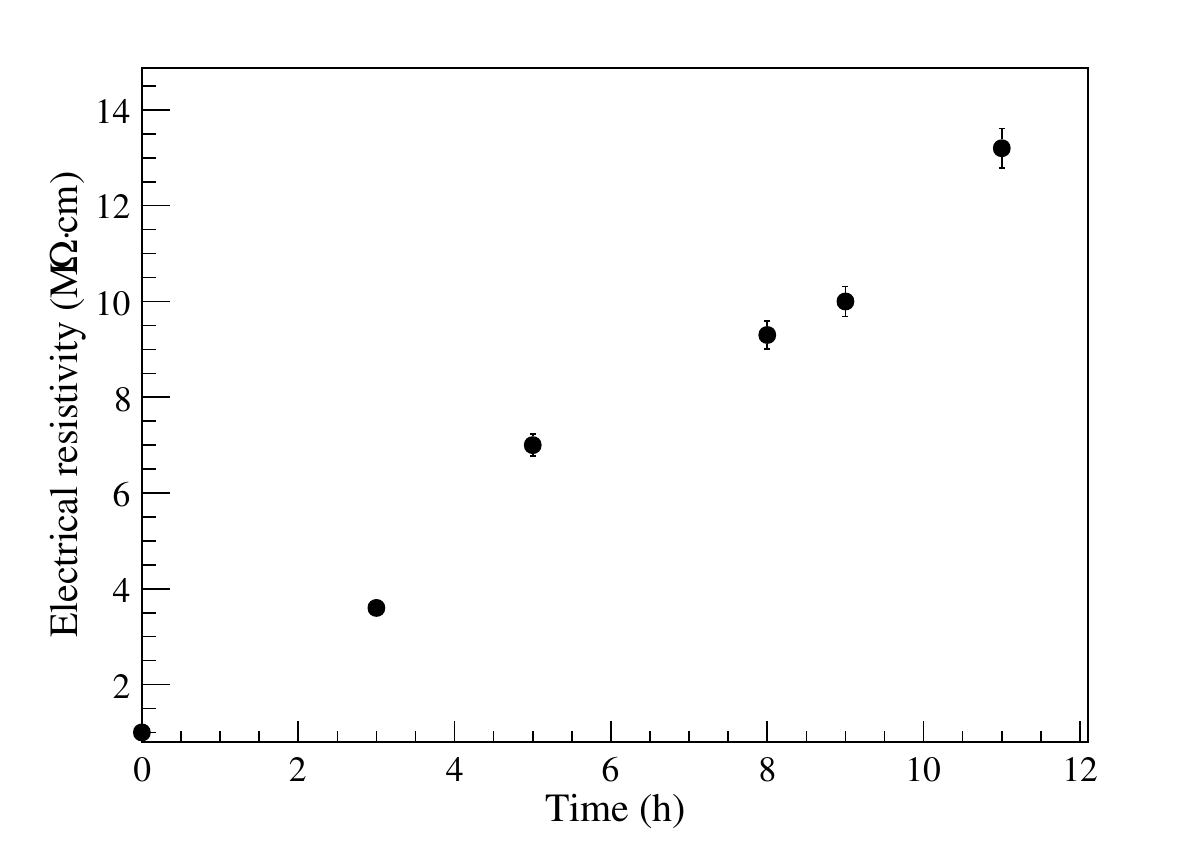}
        	\label{fig:Electrical_vs_Time}
        }
        \quad
        \subfigure[]{
            \includegraphics[width=0.4\hsize]{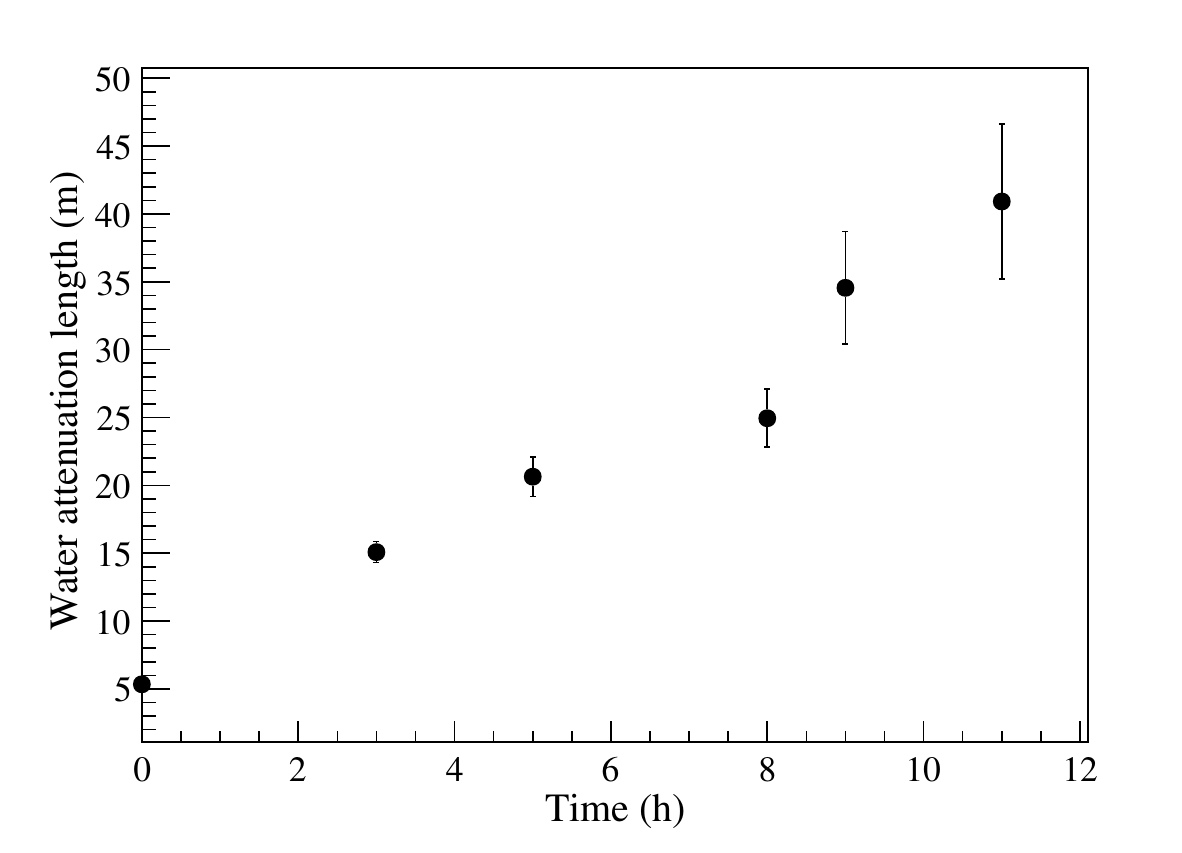}
        	\label{fig:WAL_vs_Time}
        }
        \caption{(a) Variation of water resistivity with the operating time of the circulation system. (b) Variation of the WAL with the operating time of the circulation system.}
    	\label{fig:circu_circu_r}
\end{figure}

To investigate the relationship between WAL and water quality, a water purification circulation system was employed \cite{circulation-system}. Resistivity and WAL measurements were recorded at regular intervals throughout the purification process. As depicted in Fig.~\ref{fig:Electrical_vs_Time}, resistivity increased with system operation time due to the removal of impurities and ions. Correspondingly, Fig.~\ref{fig:WAL_vs_Time} demonstrates a systematic increase in WAL with operation time, consistent with reduced light absorption and scattering from diminishing particulate matter. These measurements confirm a clear positive correlation between water quality and WAL.

\begin{figure}[htbp]
  \centering
  \includegraphics[width=0.4\textwidth]{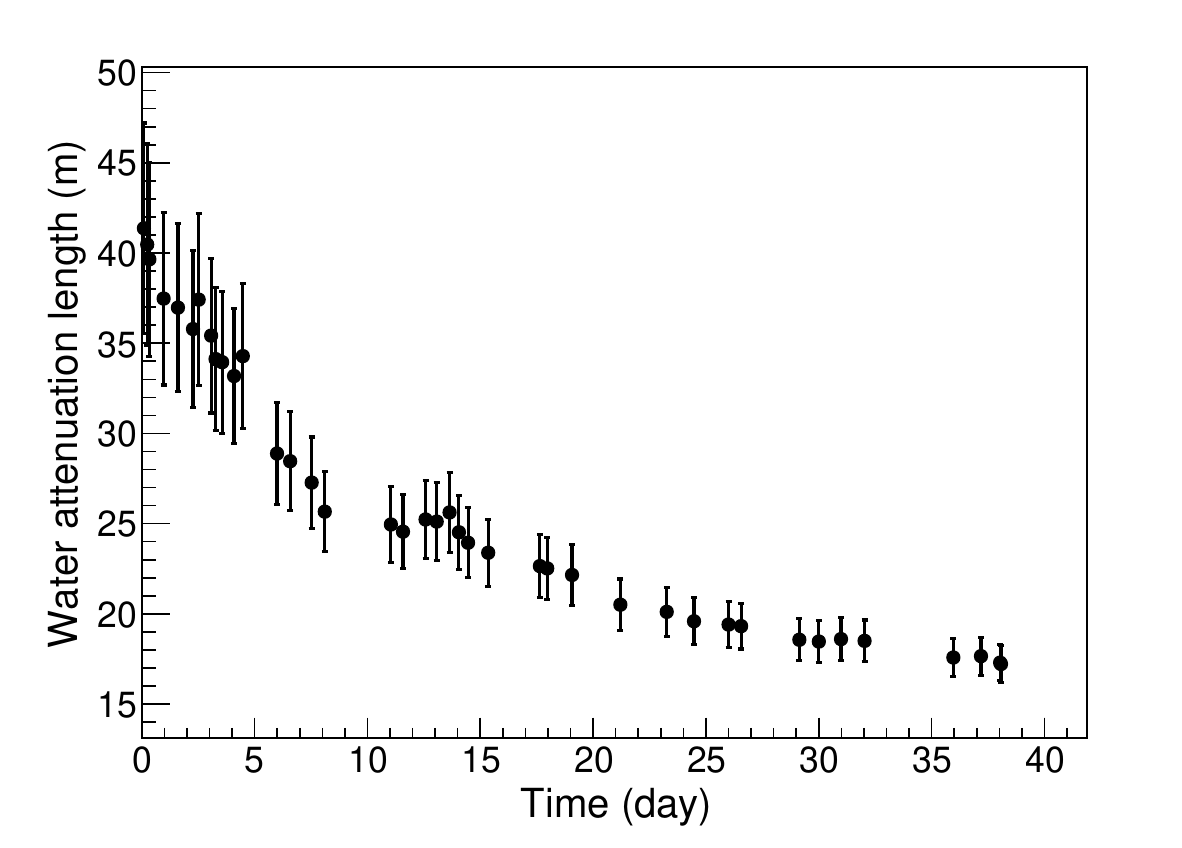}
  \caption{Long-term observation results of the WAL in the Tyvek configuration.}
  \label{fig:Tyvek}
\end{figure}   

\begin{figure}[htbp]
        \centering
        \subfigure[]{
            \includegraphics[width=0.4\hsize]{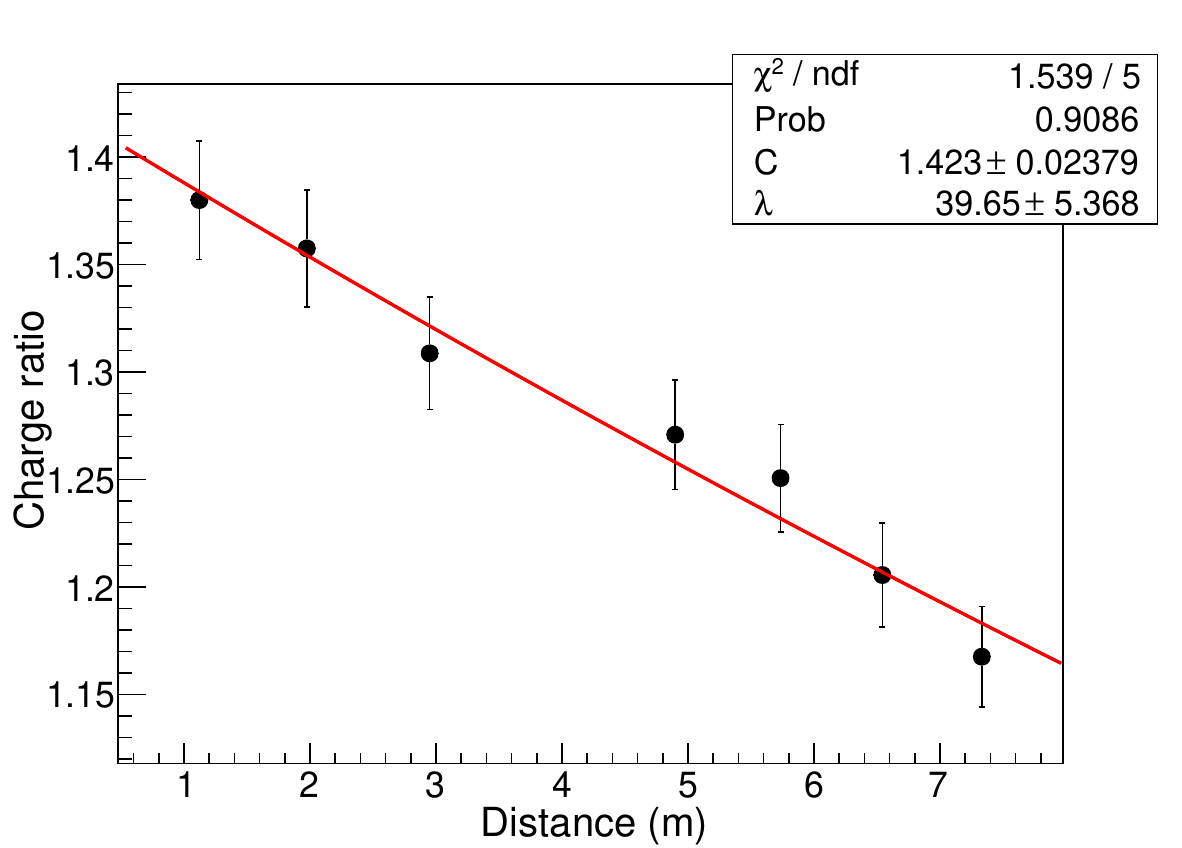}
        	\label{fig:Fit_real}
        }
        \quad
        \subfigure[]{
            \includegraphics[width=0.4\hsize]{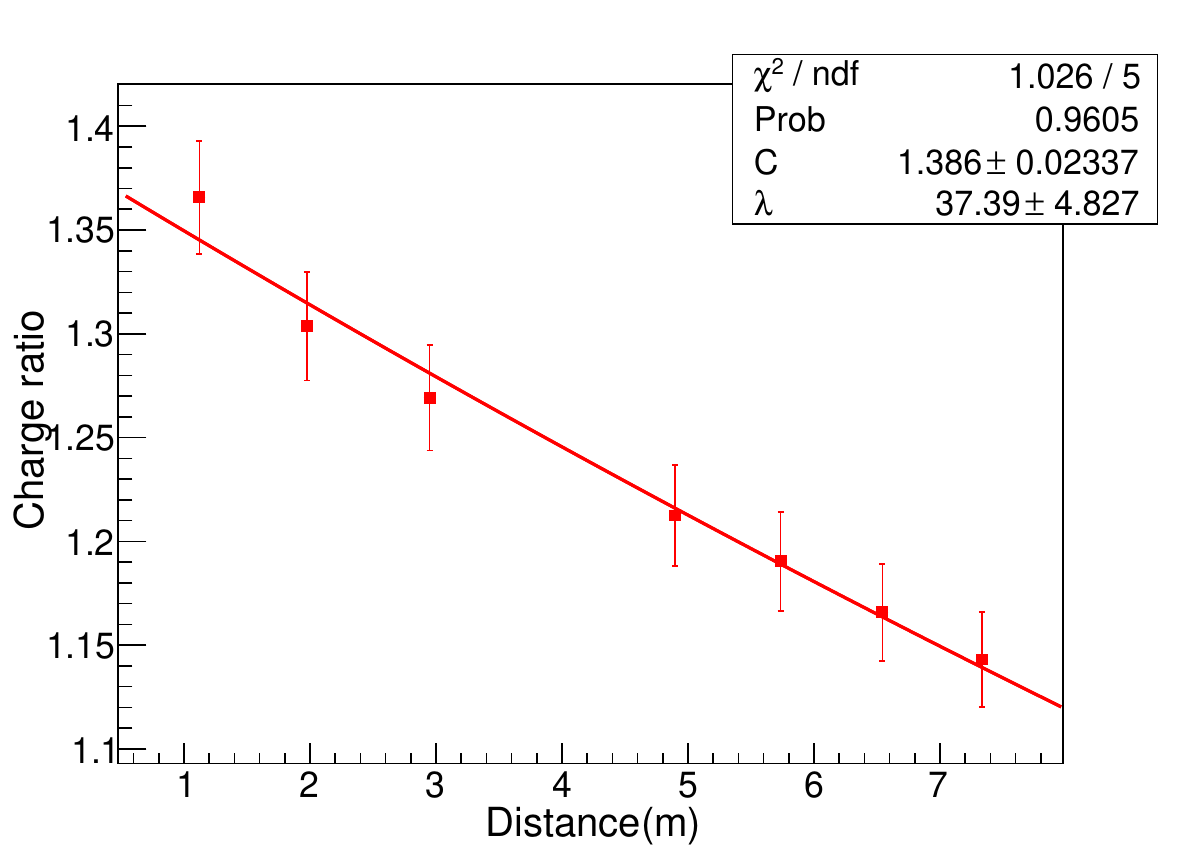}
        	\label{fig:MC}
        }
        \caption{(a) Fitting results of the measurement data for the Tyvek configuration. (b) Fitting results from the ToyMC simulation.}
        \label{fig:MC_and_real}
\end{figure}

Following the purification process, a 35-day real-time observation was conducted in the Tyvek-lined configuration without active water circulation. The observed WAL trend was similar to that of the HDPE case but demonstrated a significantly slower attenuation rate, as shown in Fig.~\ref{fig:Tyvek}. This result confirms that purification effectively removes impurities, enhances initial water quality, and delays subsequent degradation. 
Figure~\ref{fig:Fit_real} shows an example of WAL measurement and the corresponding fitting result obtained in the Tyvek configuration on the first day after water circulation was halted. The WAL was ($39.7 \pm 5.4$) m. Seven PMTs were used and the fourth PMT was excluded from the data analysis, because when opened the water tank, it was observed that the Tyvek lining above the position of the fourth PMT had partially detached from the top lid, resulting in slight sagging that obstructed the photon collection of that PMT. 

In parallel, the primary sources of measurement uncertainty were analyzed using Toy Monte Carlo (ToyMC) simulation within this 8-meter-long tank. The ToyMC is similar with the proposal of WAL measurement in 30-meter-long device in \cite{design-proposal}.
The charge measured by the PMT closest to the light source from an experimental dataset showed in Fig.~\ref{fig:Fit_real} was selected as the reference value. Based on this reference, the charges expected at the other PMTs were calculated using the inverse-square law and an assumed WAL of 39.7 meters. Variations among detector channels, photoelectron statistical fluctuations, and other relevant factors were incorporated into the simulation with about 3\% system uncertainty. The ToyMC simulation yielded a WAL value of ($37.4 \pm 4.8$) m showed in Fig.~\ref{fig:MC}, which is consistent with the experimentally measured value of ($39.7 \pm 5.4$) m showed in Fig.~\ref{fig:Fit_real}. This indicates that the fitting uncertainty from the measurement is reasonable in this prototype device, and is primarily due to the constrained physical dimensions of the apparatus. Scaling studies~\cite{design-proposal} suggest that extending the device length to 30~m would enable WAL measurements up to 100 m with an estimated uncertainty of approximately 8.1\%.

\section{Conclusion}
\label{sec5}
This study developed a compact device for WAL measurement in the laboratory, utilizing standard components such as PMTs, LEDs, and optical fibers. It is cheap, static, and easy operation for real-time water quality monitoring. The device is capable of measuring the WAL up to 50 meters. %capable of online operation in two configurations. WAL values of $48.2 \pm 8.1$ m and $41.4 \pm 5.8$ m were obtained for the HDPE-lined and Tyvek-lined configurations, respectively. 

Building upon the initial design concept, the device underwent several key optimizations: first, an ABS guide structure was implemented to mitigate the stray light effect; second, a water circulation system was integrated to enhance the purification of ultra-pure water. 
To suppress stray light, several strategies were implemented: surface roughening of the HDPE lining, installation of light-blocking baffles, and the mounting of shutters and covers on the PMTs. These measures ensured that the light intensity from the point source adhered to the inverse-square law upon incidence on the PMT cathodes. 

To further verify the accuracy of the measurements, ToyMC simulations were conducted under conditions matching the device dimensions, and the simulation results showed good agreement in 6\% with the experimental data. The 8-meter device demonstrated that a large-scale system with a 30-meter PMT separation distance can measure the WAL of up to 100 meters with an uncertainty of 8\%. The successful development and validation of this device not only confirm the feasibility of the original design concept but also provide a valuable technical reference and experimental foundation for real-time water quality monitoring in future large-scale water Cherenkov detectors.

\section{Acknowledgments}

Many thanks for the financial support from the Xie Jialin Foundation of Institute of High Energy Physics (IHEP, No.\,E3546EU2), National Natural Science Foundation of China (Grant No.\,11875282), the State Key Laboratory of Particle Detection and Electronics (No.\,SKLPDE-ZZ-202208), and the Program of Bagui Scholars Program (XF).

%references

%%%%%%%%%%%%%%%%%%%%%%%%%%%%%%

\end{document}